\documentclass[12pt,a4paper]{article}
\usepackage{graphicx}
\usepackage{cite}
\usepackage{amsmath,amssymb}
\usepackage{color}
\begin{document}
\textwidth=135mm
 \textheight=200mm
\begin{center}
{\bfseries Description of resonant states in the shell model}%
\footnote{Submitted for publication in Physics of Particles and Nuclei, http://www.maik.ru/en/journal/physpart/.}
\vskip 5mm
I.~A.~Mazur$^{\dag}$, A.~M.~Shirokov$^{\ddag, \dag, *}$, A.~I.~Mazur$^\dag$ and J.~P.~Vary$^*$ 
\vskip 5mm

{\small {\it $^\dag$ Pacific National University, Khabarovsk, Russia}} \\
{\small {\it $^\ddag$ Skobeltsyn Institute of Nuclear Physics, Lomonosov Moscow State University, Moscow, Russia}} \\
{\small {\it $^*$ Department of Physics and Astronomy, Iowa State University, Ames, IA, USA}} \\
\end{center}

\vskip 5mm

\centerline{\bf Abstract}
A technique for describing scattering states within the nuclear shell model is proposed. This technique is applied to
scattering of nucleons by $\alpha$ particles based on {\em ab initio} No-Core Shell Model calculations of $^{5}$He and $^{5}$Li nuclei
with JISP16 $NN$ interaction.
\vskip 10mm

\section{Introduction}
There is noticeable progress in the microscopic description of nuclear reactions, continuum spectra and widths of nuclear resonant states.
In particular, we mention the Lorentz Integral Transform method~\cite{Efros, LeidOrl} which was mainly utilized within
the Hyperspherical Harmonics approach and was generalized~\cite{PNavr} for the use in combination with the No-Core Shell Model (NCSM),
the  Continuum Shell Model~\cite{Volya}, the first attempts to study scattering of nucleons by nuclei within the Quantum Monte Carlo 
approach~\cite{PieWir}, the Gamow  Shell Model including the {\em ab initio} No-Core Gamow  Shell Model (NCGSM)~\cite{Jimmy}.
The main achievement in modern {\em ab initio} theory of nuclear reactions is a description of various reactions with light nuclei
within a combination of NCSM with Resonating Group Method (see reviews~\cite{LeidOrl, NavrJPG09, Vary2013}.

In this contribution, we formulate a simple method for calculating low-energy phase shifts and for 
extracting resonant energies~$E_{r}$ and widths~$\Gamma$ directly from the shell model eigenstates 
without additional complexities like introducing additional Berggren basis states as in NCGSM or additional RGM calculations as in the 
combined NCSM/RGM approach. The method is based on the $J$-matrix formalism in scattering theory~\cite{YaFi}. The $J$-matrix approach 
utilizes a diagonalization of the Hamiltonian in one of two bases: the so-called Laguerre basis that is of a particular interest for atomic physics 
applications and the oscillator basis that is appropriate for nuclear physics. The version of the $J$-matrix formalism with the oscillator basis is also 
sometimes referred to as an Algebraic Version of RGM~\cite{Fill} or as a HORSE (Harmonic Oscillator Representation of Scattering Equations) 
method~\cite{Bang}~--- we use the latter nomenclature in what follows.

The proposed method is applied to the analysis of resonant states in scattering of nucleons by $\alpha$ particle
 based on {\em ab initio} calculations of $^{5}$He and $^{5}$Li nuclei within NCSM~\cite{Vary2013}
with the JISP16 $NN$ interaction~\cite{ShirokovPLB644}.

\section{HORSE and SS HORSE formalisms}
We start with a short description of the HORSE formalism in the case of scattering in a partial wave with the orbital momentum~$\ell$
in a system of two particles  with reduced mass~$\mu=\frac{m_{1}m_{2}}{m_{1}+m_{2}}$
interacting via potential~$V$. The relative motion 
wave function is expanded in infinite series of oscillator functions with the oscillator frequency~$\hbar\Omega$
 labeled  by the principal quantum
number~$n=0,1,2,...,\infty$ or by the oscillator quanta~$N=2n+\ell$.

The kinetic energy matrix in the oscillator basis is tridiagonal, its non-zero matrix
elements~$T_{N,N}$ and~$T_{N,N\pm 2}$ are increasing linearly
with~$N$. On the other hand, the potential energy matrix elements~$V_{N,N'}$ are decreasing
with~$N$ and/or~$ N'$. Therefore a reasonable approximation is to neglect  
the potential energy matrix elements~$V_{N,N'}$ with respect to~$T_{N,N'}$ 
if~$N$ or~$N'>{\cal N}$. In other words, we split the complete infinite oscillator basis space into two
subspaces: the `internal' subspace~$P$ spanned by oscillator functions with~$N\leq {\cal N}$ where the
complete Hamiltonian~$H=T+V$ is used and `external' subspace~$Q$ spanned by oscillator functions 
with~$N > {\cal N}$ corresponding to the free motion where the Hamiltonian includes only the kinetic energy.

The eigenvectors of the infinite Hamiltonian matrix including only kinetic energy matrix elements in the $Q$ space and both potential
and kinetic energy matrix elements in the $P$ space can be found if the eigenenergies~$E_\nu$ and 
eigenvectors~$\langle N|\nu\rangle$ of the Hamiltonian submatrix in the $P$ space are known.  $E_\nu$ 
and~$\langle N|\nu\rangle$ fit the set of linear equations
\begin{gather}
\sum_{n'=0}^{d-1} 
 H_{N,2n'+\ell}\;\langle 2n'+\ell |\nu\rangle  =E_{\nu}\langle N|\nu\rangle, \quad N\leq {\cal N},
         \quad \nu=0,1,...,d-1.
\label{Eigen}
\end{gather}
Here $d=({\cal N}-\ell)/2+1$ 
is the dimensionality of the $P$ space.
All scattering observables at any energy~$E$ can be extracted from the eigenvectors of this infinite Hamiltonian matrix. For example,
the scattering phase 
 shifts~$\delta_\ell(E)$ can be calculated as~\cite{Bang}
\begin{equation}
	\label{J_phase}
	\tan\delta_\ell(E)=-\frac{S_{{\cal N},\ell}(E)-G_{{\cal N},\,{\cal N}}(E)\:T_{{\cal N},\,{\cal N}+2}\:S_{{\cal N}+2,\ell}(E)}{C_{{\cal N},\ell}(E)-G_{{\cal N},\,{\cal N}}(E)\:T_{{\cal N},\,{\cal N}+2}\:C_{{\cal N}+2,\ell}(E)},
\end{equation}
where
\begin{equation}
	\label{JGnn}
G_{{\cal N},{\cal N}}(E)=-\sum_{\nu=0}^{d-1} 
 \frac{\left|\left<{\cal N}|\nu\right>\right|^2}{E_\nu-E},
\end{equation}
and~$S_{N,\ell}(E)$ and~$C_{N,\ell}(E)$ are   the regular and  irregular solutions of the free Hamiltonian in the oscillator representation which analytical expressions can be found in Ref.~\cite{Bang}.

The HORSE formalism can be used in combination with any approach utilizing the oscillator basis expansion. In particular, it can be used to generalize the
nuclear shell model for applications to the continuum spectrum. In this case, the $P$ space should be
associated with the many-body shell model basis space while
the $Q$ space is to be used to `open' a particular channel in the many-body system. The standard matrix equation defining the shell model 
eigenstates should be used as the
$P$-space set of linear equations~\eqref{Eigen} where the relative motion wave function components in the oscillator 
basis~$\langle N|\nu\rangle$ should be replaced by many-body oscillator shell-model 
components~$\langle N[\alpha] J|\nu\rangle$ characterized by a given value of the total angular momentum~$J$ and an additional
set of quantum numbers~$[\alpha]$ which distinguish many-body oscillator states with the same~$N$ and~$J$. The summation
in Eq.~\eqref{Eigen} should run over all possible states with different~$[\alpha]$ thus increasing drastically the  $P$ space dimensionality~$d$.
This increase of the $P$ space dimensionality is however just a manifestation of the increase of basis space in a many-body system
and is implemented in modern shell model codes. 
A more significant limitation for applications
is the same increase of the number
of summed terms in Eq.~\eqref{JGnn} where the last oscillator components of the relative motion
eigenfunctions~$\left<{\cal N}|\nu\right>$ should be replaced by the many-body oscillator 
components~$\left<{\cal N} [\alpha] J\Gamma|\nu\right>$ with the maximal total oscillator quanta in the $P$ space~{$\cal N$} 
projected onto a desired channel~$\Gamma$. Note, Eq.~\eqref{JGnn} requires summation over all shell model eigenstates~$E_{\nu}$
with a given value of the total angular momentum~$J$ while the modern shell model codes usually are designed to calculate only
a few lowest eigenstates. The high-lying eigenstates~$E_{\nu}$ can contribute even to low-energy phase shifts since the increase
of the denominator in Eq.~\eqref{JGnn} can be accompanied for some states by an increase of the numerator. 

To overcome these difficulties, we propose a Single-State (SS) HORSE formalism. The conventional wisdom says that a shell model
eigenstate~$E_\nu$ defines all the properties of a nearby resonant state. So, let us calculate the phase shift~$\delta_\ell(E_{\nu})$
at this energy. From Eqs.~\eqref{J_phase}--\eqref{JGnn} we obtain a very simple expression:
\begin{equation}
 \label{SSJM_phase}
\tan{\delta_{\ell} (E_{\nu})}  = -\frac{S_{{\cal N}+2,\ell}(E_{\nu})}
     {C_{{\cal N}+2,\ell}(E_{\nu})} .
\end{equation}
Note, we get rid not only of the need to sum over a huge number of eigenstates as in Eq.~\eqref{JGnn} but also from the shell model
wave function component~$\langle{\cal N}[\alpha] J\Gamma|\nu\rangle$  defining the desired channel. Hence Eq.~\eqref{SSJM_phase}
can be used  for scattering channels of any type. In the  case of low-energy scattering 
when~$E\ll\frac18\hbar\Omega({\cal N}+2-\ell)^{2} $, one can use asymptotic expressions for~$S_{{\cal N}+2,\ell}(E_{\nu})$
and~$C_{{\cal N}+2,\ell}(E_{\nu})$ at large~${\cal N}$~\cite{democr} to obtain
\begin{equation} 
\label{SSJM_phase_sc}
\tan{\delta_{\ell} (E_{\nu})}  = \frac{j_\ell(2\sqrt{E_\nu/s})}{n_\ell(2\sqrt{E_\nu/s})},\qquad s = \frac{\hbar\Omega}{{\cal N}+7/2},
\end{equation}
where~$j_\ell(x)$ and~$n_\ell(x)$ are spherical Bessel and Neumann functions. Equation~\eqref{SSJM_phase_sc} exhibits a scaling
property of low-energy scattering: the phase shift~$\delta_{\ell} (E)$ at shell model eigenenergies~$E= E_{\nu}$ does not depend on the shell model
parameters~$\hbar\Omega$ and~${\cal N}$ individually but only on their combination~$s$.

The shell model calculations are usually performed for  sets of~${\cal N}$ and~$\hbar\Omega$ values. Within the SS HORSE
formalism, we can calculate the phase shift~$\delta_{\ell} (E)$ at the respective set of 
eigenenergies~$E=E_\nu({\cal N} ,\hbar\Omega)$ covering some energy interval. Next we can extrapolate  the phase shift
on a larger energy interval using accurate parametrizations of~$\delta_{\ell} (E)$ at low energies.

\section{Low-energy phase shift parametrization}
The scattering $S$-matrix as a function of momentum~$k$ is known~\cite{Taylor} to have the following symmetry property:
\begin{gather}
S_{\ell}(k)=S^{{-1}}_{\ell}(-k).
\label{SymmS}
\end{gather}
Since $S_{\ell}=e^{2i\delta_{\ell}}$, the phase shift~$\delta_{\ell}(E)$ is an odd function of~$k$ and its expansion in
Taylor series of~$\sqrt{E}\sim k$ includes only odd powers of~$\sqrt{E}$:
\begin{gather}
\delta_{\ell}(E)=c\,\sqrt{E}+d\bigl(\sqrt{E}\bigr)^{3}+...
\label{Taylordelta}
\end{gather}
More, since~$\delta_{\ell} \sim k^{2\ell+1}$ in the limit~$k\to0$, $c=0$ in the case of $p$-wave scattering, $c=d=0$ in the case of
$d$-wave scattering, etc.

If the $S$-matrix has a pole associated with a bound state at the imaginary momentum~$k=ik_b$
 or a pole associated with a low-energy resonance at the complex momentum~$k=\kappa_r\equiv k_r-i\gamma_r$,  it can be expressed as
\begin{equation} 
\label{S_struct}
   S(k)=\Theta(k)S_p(k),
\end{equation}
where $\Theta(k)$ is a smooth function of~$k$ and the pole term~$S_p(k)$ in the case of a bound state ($p=b$) or a resonant
state $(p={r})$ is~\cite{Taylor}
\begin{equation}
    S_b(k)=\frac{k+ik_b}{k-ik_b}, \qquad  S_r(k)=\frac{(k+\kappa_r)(k-\kappa^{*}_r)}{(k-\kappa_r)(k+\kappa^{*}_r)}.
    \label{SbSr}
\end{equation}
The respective phase shift
\begin{equation}
   \delta_\ell(k)=\phi(k)+\delta_p(k),
   \label{deltaphip}
\end{equation}
where the pole contribution~$\delta_{p}(k)$ takes the form
\begin{equation} 
\label{deltapE}
\delta_b(E)=\pi-\arctan\sqrt{\frac{E}{|E_b|}},
\qquad   \delta_r(E) = -\arctan
                     \frac{a\sqrt{E}}{E-b^2}. 
\end{equation}
Here $\pi$ in the expression for~$\delta_b$ appears due to the Levinson theorem~\cite{Taylor},
$E_b = - \frac{\hbar^2 k_b^2}{2\mu} <0$ is the bound state energy while the resonance energy~$E_r$ and its width~$\Gamma$ are\strut 
\begin{equation}
\label{EGab}
E_r \equiv \frac{\hbar^2}{2\mu} (k_r^2-\gamma_r^2) = b^2-a^2/2,\qquad 
\frac{\Gamma}{2} \equiv \frac{\hbar^2}{2\mu}k_r \gamma_r = a\sqrt{b^2-a^2/4}.
\end{equation}

In applications to the non-resonant $n\alpha$ scattering in the~$\frac{1}{2}^{+}$ state, we therefore are using the following
parametrization of the phase shift:
\begin{equation} 
\label{S_phase_b}
   \delta_0(E)=\pi-\arctan\sqrt{\frac{E}{|E_b|}}+c\sqrt{E}+d\bigl(\sqrt{E}\bigr)^3.
\end{equation} 
The bound state pole contribution here is associated with the so-called Pauli-forbidden state.
There are resonances in the $n\alpha$ scattering in the~$\frac{1}{2}^{-}$ and~$\frac{3}{2}^{-}$ states; hence we parametrize the
phase shifts as
\begin{equation} 
\label{S_phase_r}
   \delta_1(E)=-\arctan
                   \frac{a\sqrt{E}}{E-b^2}
                                         -\frac{a}{b^2}\sqrt{E} + d\bigl(\sqrt{E}\bigr)^3.
\end{equation}
This form guarantees that $ \delta_1\sim k^{3}$ in the limit of~$E\to0$ [see Eq.~\eqref{Taylordelta} and discussion below it].

\section{Application to $N\alpha$ scattering}

We calculate the $N\alpha$ scattering phase shifts and resonant parameters using the results of the NCSM calculations of $^{5}$He
and $^{5}$Li nuclei with the JISP16 $NN$ interaction. However, we should note here that we first carefully verified 
the computational algorithm described below supposing $\alpha$ as a structureless particle and using phenomenological
$N\alpha$ potentials. In this case, the scattering phase shifts and resonant pole locations can be calculated numerically. Our SS HORSE
approach was found to be very accurate.

The NCSM model space is conventionally truncated using~$N_{\max}$, the maximal excitation oscillator quanta. This NCSM
model space should be associated with the $P$ space of the SS HORSE method which is defined using total oscillator quanta in
the many-body system, ${\cal N}$, which is entering the above SS HORSE formulas.
In the case of $^{5}$He and $^{5}$Li nuclei, we set~${\cal N}=N_{\max}+1$. Note, even~$N_{\max}$ values 
should be used to calculate
the  natural parity states~$\frac{1}{2}^{-}$ and~$\frac{3}{2}^{-}$  in these nuclei while the unnatural parity state~$\frac{1}{2}^{+}$ is obtained
in the NCSM calculations with odd~$N_{\max}$ values.
In particular, we perform here the NCSM calculations with~$N_{\max}=2$, 4,~...\,, 16 for~$\frac{1}{2}^{-}$ and~$\frac{3}{2}^{-}$ states and
with~$N_{\max}=1$, 3,~...\,, 15 for the~$\frac{1}{2}^{+}$ state.
We pick up for further scattering calculations the lowest NCSM eigenenergies~$E_{0}^{\rm NCSM}$
in $^{5}$He and $^{5}$Li  with~$J^{\pi}=\frac{3}{2}^{-}$, $\frac{1}{2}^{-}$ and~$\frac{1}{2}^{+}$; note, all these~$E_{0}^{\rm NCSM}<0$
since they are defined regarding to the 5-nucleon decay threshold. The SS HORSE method requires however positive
eigenenergies~$E_{0}$ defined in respect to the~$N+\alpha$ threshold. We obtain these eigenenergies 
as~$E_{0}=E_{0}^{\rm NCSM}-E_{0}^{\alpha}$ where  $E_{0}^{\alpha}$ is the $^{4}$He ground state energy obtained in NCSM with
the same~$\hbar\Omega$ and the same~$N_{\max}$ in the case of~$\frac{1}{2}^{-}$ and~$\frac{3}{2}^{-}$ states and with
excitation quanta~$N_{\max}-1$ in the case of unnatural parity $\frac{1}{2}^{+}$  states of $^{5}$He and $^{5}$Li.

\begin{figure}[t!]
\centerline{\includegraphics[width=0.477\textwidth]{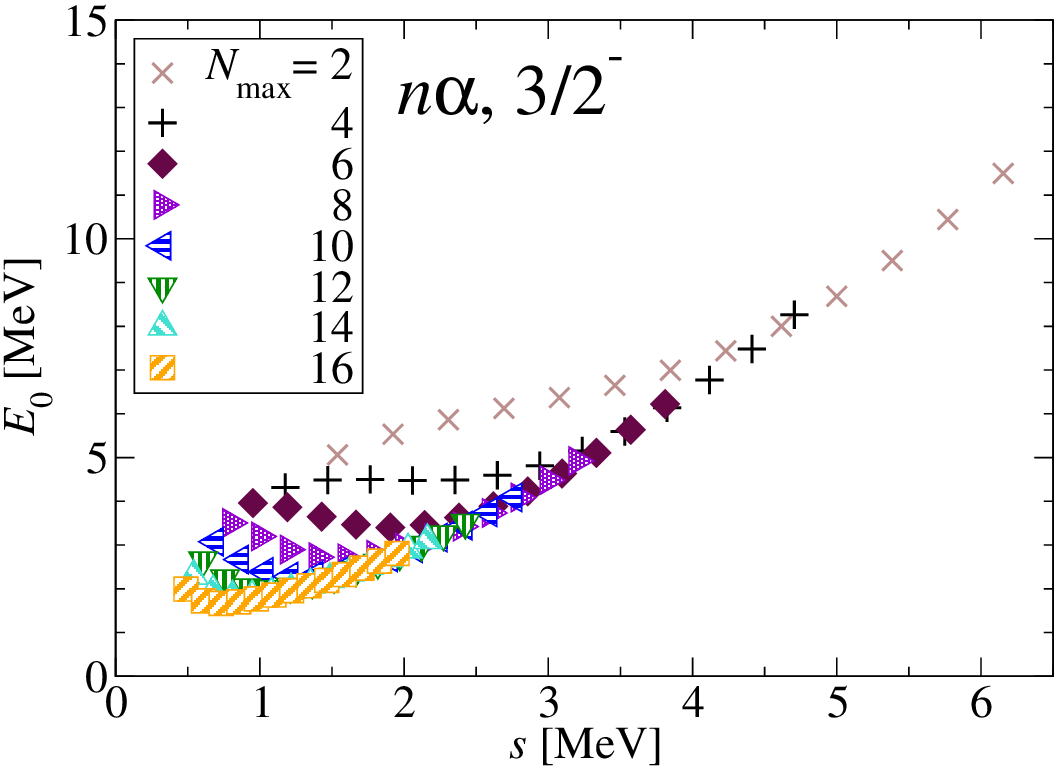} \hfill \includegraphics[width=0.5\textwidth]{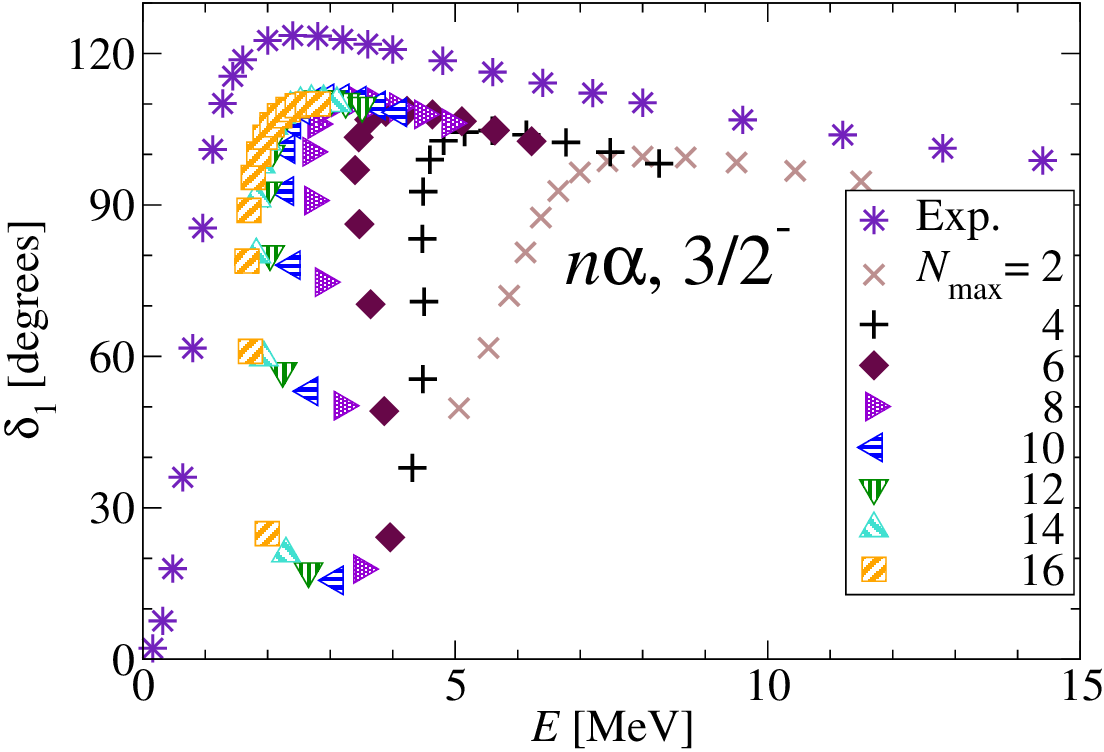}}\vskip 3mm
\centerline{\includegraphics[width=0.478\textwidth]{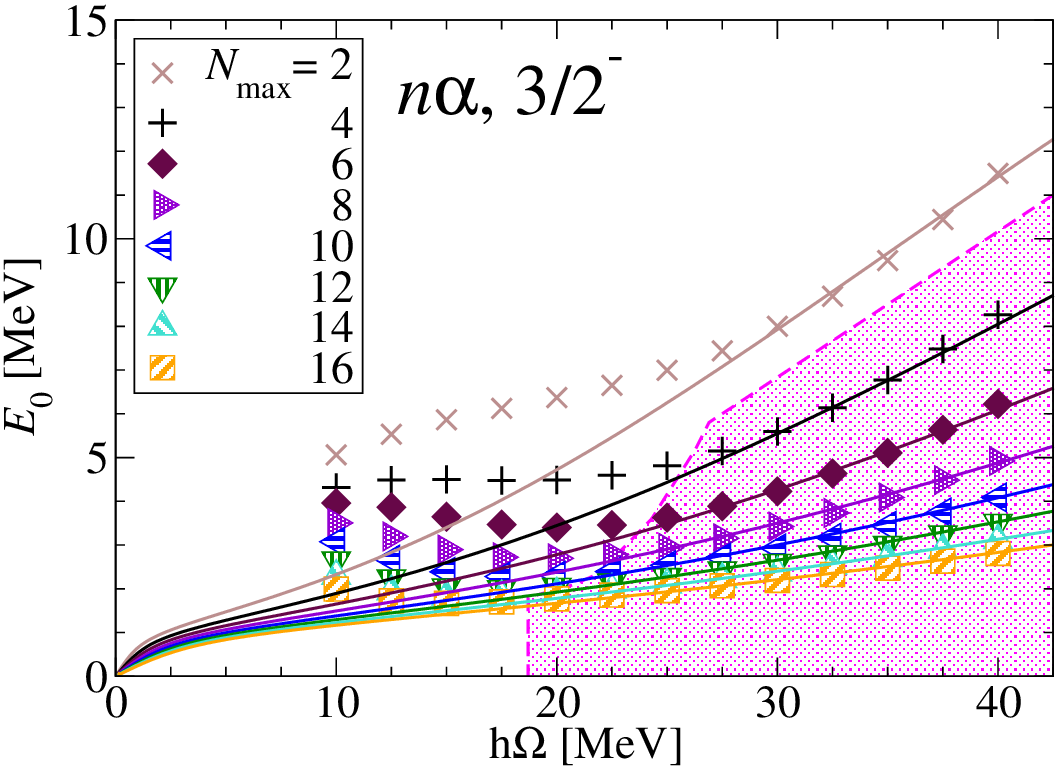}\hfill \includegraphics[width=0.5\textwidth]{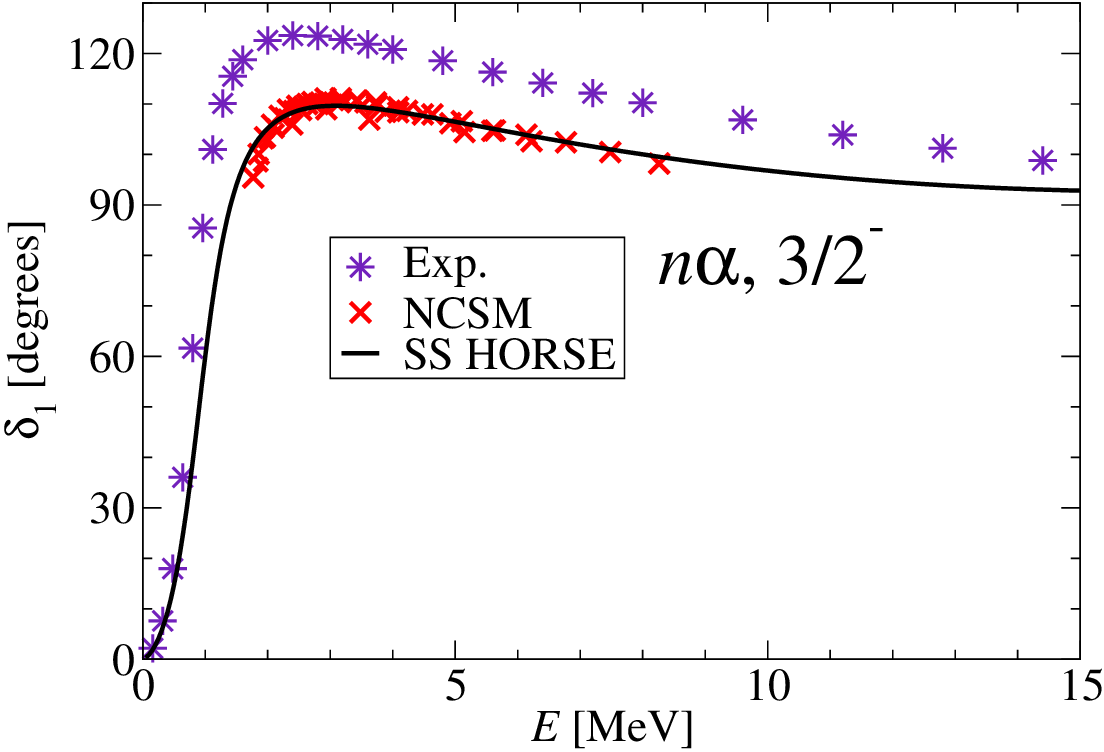}}
\caption{$n\alpha$ scattering in the~$\frac{3}{2}^{-}$ state. Left:
eigenenergies~$E_{0}$ obtained with various~$N_{\max}$ vs scaling parameter~$s$ (upper panel) and vs~$\hbar\Omega$ (lower panel).
The shaded area shows the~$E_{0}$ values selected for the SS HORSE analysis. The lines were obtained by Eq.~\eqref{ImplicitENhw} with
fitted parameters. Right: the phase shift~$\delta_{1}(E)$ obtained by Eq.~\eqref{SSJM_phase} vs the $n\alpha$ 
c.\:m. energy. The symbols on the upper panels shows the phase shifts obtained from all~$E_{0}$ values while the lower panel depicts the 
phase shifts generated by the selected~$E_{0}$ values. The line was obtained by Eq.~\eqref{S_phase_r} with fitted parameters. The 
experimental phase
shifts are taken from Ref.~\cite{BondNPA77}. }
\label{n-4Hep32_sc}      
\end{figure}

After defining eigenenergies~$E_{0}$, we note that not all of them can be used for phase shift calculations due to convergence patterns
of eigenstates in the harmonic oscillator basis. Our SS HORSE formalism results in relations for phase shifts similar to those obtained in
Ref.~\cite{More}. Using the nomenclature of Ref.~\cite{More}, we should use only eigenenergies~$E_{0}$ which are not influenced 
by infra-red corrections. As an example, we discuss the selection of eigenenergies~$E_{0}$ in the case of $n\alpha$ scattering
in the~~$\frac{3}{2}^{-}$ state. The $^{5}$He calculations were performed with~$\hbar\Omega$ ranging
from~10 to~40~MeV in steps of~2.5~MeV and, as was mentioned above, with~$N_{\max}\leq 16$
using the code MFDn~\cite{Maris_2010_2}.
The obtained~$E_{0}$ values are depicted in the left panels of Fig.~\ref{n-4Hep32_sc}. Due
to the scaling property~\eqref{SSJM_phase_sc}, we expect  all eigenenergies~$E_{0}$ as function of the scaling parameter~$s$ 
to lie on a single curve. We see however deviations from such a curve on the left upper panel of Fig.~\ref{n-4Hep32_sc} that occur
for each set of~$E_{0}$ obtained with a given~$N_{\max}$ below some critical~$\hbar\Omega$ value. This critical~~$\hbar\Omega$ value
decreases with increasing~$N_{\max}$. More instructive are the  phase shifts~$\delta_{1} (E_{0})$ obtained by Eq.~\eqref{SSJM_phase}
which are also expected to form a single curve. The deviations from this curve are seen in the upper right panel of Fig.~\ref{n-4Hep32_sc} to be
more pronounced. For the calculation of the phase shifts and resonant parameters, we select the~$E_{0}$ values which form approximately
single curves on upper panels of Fig.~\ref{n-4Hep32_sc}. This selection is illustrated by lower panels of Fig.~\ref{n-4Hep32_sc}.

The resonant $n\alpha$ scattering phase shifts in the~$\frac{3}{2}^{-}$ and~$\frac{1}{2}^{-}$ states are described by Eq.~~\eqref{S_phase_r}. We
need to fit the parameters~$a$, $b$ and~$d$ of this equation.
The resonance energy~$E_{r}$ and width~$\Gamma$
can then be obtained by Eq.~\eqref{EGab}. From Eqs.~\eqref{SSJM_phase} and~\eqref{S_phase_r} we derive  the following relation for resonant $n\alpha$ 
scattering in the~$\frac{3}{2}^{-}$ and~$\frac{1}{2}^{-}$ states:
\begin{equation} 
\label{ImplicitENhw}
   -\frac{S_{N_{\max}+3,\,1}(E_{0})}{C_{N_{\max}+3,\,1}(E_{0})} = 
   \tan\!\left(\!-\arctan\frac{a\sqrt{E_0}}{E_0-b^2}-\frac{a}{b^2}\sqrt{E_0}+d\bigl(\sqrt{E_0}\bigr)^3 \!\right)\!.
\end{equation}
We assign some values to the parameters~$a$, $b$ and~$d$ and solve this equation to find~$E_{0}$ for each desired combination
of~$N_{\max}$ and~$\hbar\Omega$ values (note, $\hbar\Omega$ enters definition of functions~$S_{N,\ell}(E)$ 
and~$C_{N,\ell}(E)$~--- see, e.\:g., 
Ref.~\cite{Bang}). The resulting set of~$E_{0}$ is compared with the set of selected eigenvalues obtained from NCSM and we
minimize the rms deviation between these two sets to find the optimal values of the parameters~$a$, $b$ and~$d$. The behavior
of~$E_{0}$ as functions of~$\hbar\Omega$ dictated by Eq.~\eqref{ImplicitENhw} with the fitted optimal parameters~$a$, $b$ and~$d$ for 
various~$N_{\max}$ values is depicted by curves on the lower left panel of Fig.~\ref{n-4Hep32_sc}. It is seen that these curves accurately
describe the selected eigenvalues from the shaded area. The phase shifts~$\delta_{1}(E)$ obtained by Eq.~\eqref{S_phase_r} with 
fitted parameters are shown in the lower right panel of Fig.~\ref{n-4Hep32_sc}. It is seen that our theoretical predictions are in a reasonable
correspondence with the results of phase shift analysis of experimental scattering data of Ref.~\cite{BondNPA77}.

\begin{figure}
\centerline{\includegraphics[width=0.49\textwidth]{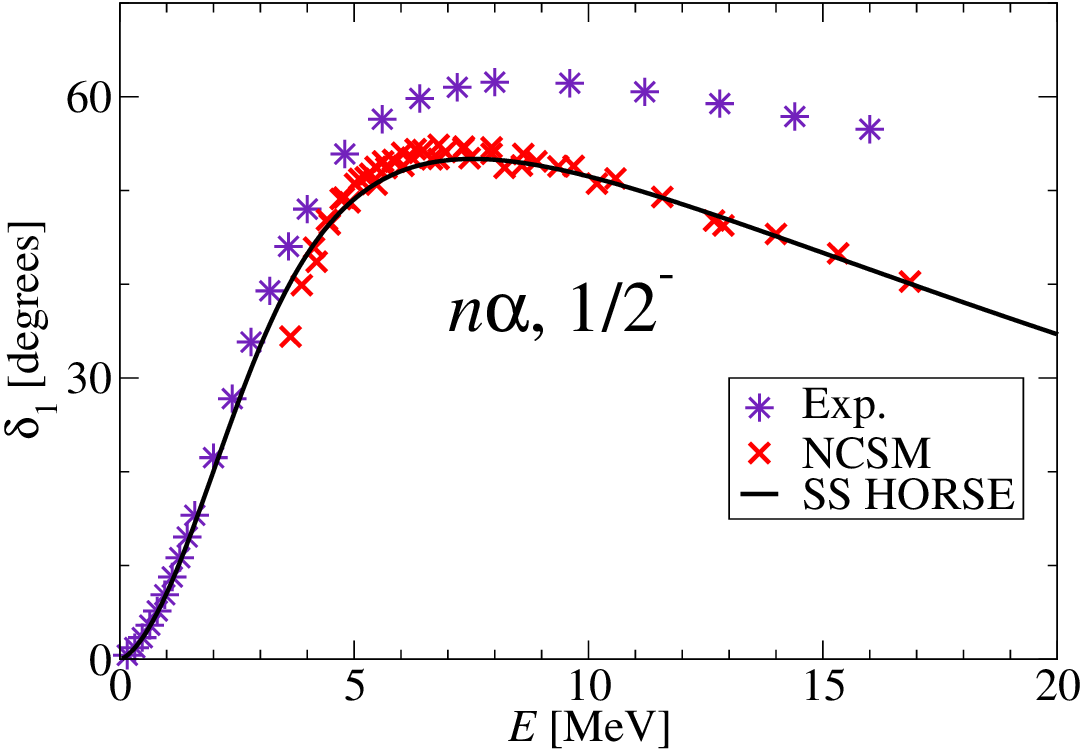}\hfill \includegraphics[width=0.5\textwidth]{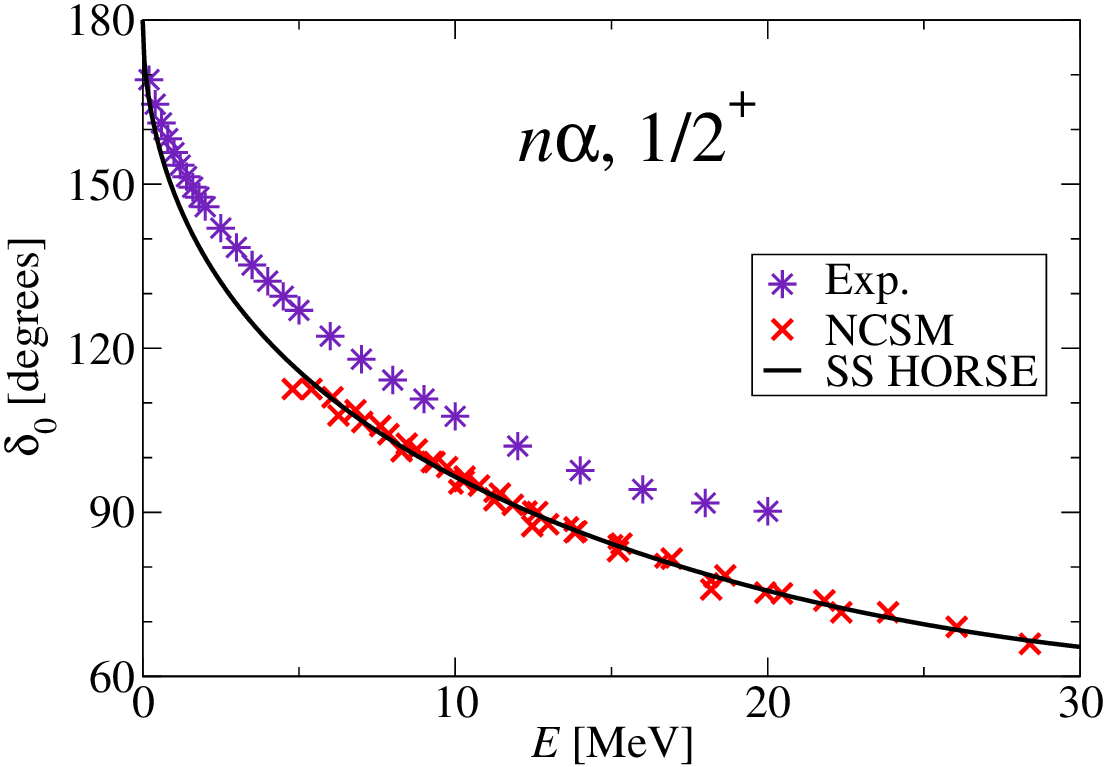}}
\caption{$n\alpha$ scattering phase shifts in the~$\frac{1}{2}^{-}$ (left) and~$\frac{1}{2}^{+}$ (right) states.  See Fig.~\ref{n-4Hep32_sc}
for details.}
\label{n-4Hep12s12_Select1}      
\end{figure}

A wider $\frac{1}{2}^{-}$ resonance and non-resonant $n\alpha$ scattering phase shifts in the~$\frac{1}{2}^{+}$ state are described in the same 
manner (see Fig.~\ref{n-4Hep12s12_Select1}).  The only difference in the case of the~$\frac{1}{2}^{+}$ scattering is that instead of
Eq.~\eqref{ImplicitENhw}, we are using  
\begin{equation} \label{ImplicitENhw_s}
    -\frac{S_{N_{\max}+3,\,0}(E_{0})}{C_{N_{\max}+3,\,0}(E_{0})} = 
    \tan\!\left(\!\pi-\arctan\sqrt{\frac{E_\nu}{|E_b|}}+c\sqrt{E_\nu}+d\bigl(\sqrt{E_0}\bigr)^3 \!\right)\! ,
\end{equation}
which can be easily obtained from Eqs.~\eqref{SSJM_phase} and~\eqref{S_phase_b}. The phase shift analysis of experimental data
is also reasonably described in these cases.

The formalism presented in Refs.~\cite{Bang, PRCnal} can be used to generalize the SS HORSE approach to charged
particle scattering. This generalization yields a more complicated formula for the SS HORSE phase shifts than Eq.~\eqref{SSJM_phase} and
to other relations derived from it like Eqs.~\eqref{ImplicitENhw} and~\eqref{ImplicitENhw_s}. However, a modified
scaling property~\eqref{SSJM_phase_sc} can also be obtained
in this case and we can use generally the same fitting algorithm for the parameters describing the phase shifts.

\begin{table}
\caption{Energies~$E_{r}$ and widths~$\Gamma$ (in MeV) of  $^5$He and $^5$Li resonant states.}
\label{N-4Hep32p12}  
\begin{center}
\begin{tabular}{c|cccc|cccc}
& \multicolumn{2}{c}{$^5$He$(\frac32^-)$} & \multicolumn{2}{c|}{$^5$He$(\frac12^-)$} & \multicolumn{2}{c}{$^5$Li$(\frac32^-)$} & \multicolumn{2}{c}{$^5$Li$(\frac12^-)$} \\
 & $E_r$ & $\Gamma$ & $E_r$ & $\Gamma$ & $E_r$ & $\Gamma$ & $E_r$ & $\Gamma$ \\
\hline
$N_{\max}=4\div16$ & 0.93 & 1.01 & 1.84 & 5.49 & 2.05 & 1.35 & 3.29 & 4.70 \\
$N_{\max}=4\div6$ & 0.97 & 1.07 & 1.82 & 5.61 & 2.72 & 1.27 & 3.83 & 4.57 \\
\hline
 $R$-matrix \cite{CsotoPRC97} & 0.80 & 0.65 & 2.07 & 5.57 & 1.69 & 1.23 & 3.18 & 6.60 \\
\end{tabular}
\end{center}
\end{table}

Resonance energies~$E_{r}$ and widths~$\Gamma$ obtained using Eq.~\eqref{EGab} are presented in Table~\ref{N-4Hep32p12}. We
show in the Table not only the results obtained from NCSM calculations with~$N_{\max}$ ranging from~4 to~16 but also from
calculations with~$N_{\max}\le6$ which demonstrate that the resonant parameters only slightly change when the fit is performed using
the NCSM results restricted to an essentially smaller model space. This is very encouraging for future applications to heavier nuclear
systems. Our results are in a good agreement with $R$-matrix analysis of experimental data of Ref.~\cite{CsotoPRC97}.\vskip 2mm

{\em Acknowledgements}. This work was supported in part by the Russian Foundation for Basic Research
under Grants No.~15-32-51239 and~15-02-06604, by the Ministry of Education and Science of Russian Federation,
by the US Department of Energy under Grants No. DESC0008485 (SciDAC/NUCLEI) and 
No.~DE-FG02-87ER40371 and by the US National Science Foundation under Grant No. 0904782.
Computational resources were provided by the National Energy Research Supercomputer Center (NERSC), which is supported by the Office of Science of the U.S. Department of Energy under Contract No. DE-AC02-05CH11231.

\mbox{}\\

\end{document}